# On the $^{12}$C/$^{13}$C carbon isotope effect in the quasi one-dimensional superconductor Sc$_3$CoC$_4$

Christof D. Haas,[1] Georg Eickerling,[1] Ernst-Wilhelm Scheidt,[1] Dominik Schmitz,[1] Jan Langmann,[1] Jian Lyu,[2] Junying Shen,[2] Rolf Lortz,[2] Daniel Eklöf,[3] Jan Gerrit Schiffmann,[1] Leo van Wüllen,[1] Anton Jesche,[1] Wolfgang Scherer.[1]

[1] Institut für Physik, Universität Augsburg, Universitätsstraße 1, 86135 Augsburg, Germany.
[2] Department of Physics and the William Mong Institute of Nano Science and Technology, Hong Kong University of Science and Technology, Clear Water Bay, Kowloon, Hong Kong, China.
[3] Department of Materials and Environmental Chemistry, Stockholm University, S-10691 Stockholm, Sweden.

Sc$_3$CoC$_4$ is the only superconductor in the group of the isotypic carbides Sc$_3$MC$_4$ (M = Mn, Fe, Ru, Os, Co, Rh, Ir, Ni), rendering it into an ideal benchmark system to systematically study the prerequisites and mechanism of superconductivity in such quasi one-dimensional structures. To investigate the isotope effect, the substitution series Sc$_3$Co($^{12}$C$_{1-x}$$^{13}$C$_x$)$_4$ with $x$ = 0, 0.5 and 1 was synthesized by arc melting. The sample homogeneity was confirmed by powder X-ray diffraction and NMR spectroscopy. The resulting isotope coefficient based on magnetization studies on polycrystalline samples and electrical resistivity measurements on single crystals of $\alpha$ = 0.58 lies close to the value of 0.5 predicted by BCS theory.




**Introduction**

The transition metal carbide $Sc_3CoC_4$ is considered as benchmark system for the phenomenon of superconductivity in compounds featuring quasi one-dimensional (quasi 1D) structural motives [1,2,3,4]. The exploration of its low-temperature physical properties in 2010 revealed a superconducting phase transition at $T_c$ = 4.5 K and a Peierls-type structural transformation from an orthorhombic (*Immm*) high-temperature (HT) to a monoclinic (*C2/m*) low temperature (LT) phase below 143 K [1,5,6]. In the HT modification, $Sc_3CoC_4$ contains quasi 1D $[CoC_4]_\infty$ ribbons which are embedded in a matrix of scandium atoms (Figure 1a,b). However, detailed experimental electron density studies and band structure analyses reveal a rather complex electronic structure characterized by strong covalent Co-C bonds and significant interactions between the $C_2$ dicarbido moieties and the surrounding scandium matrix [7]. At low temperatures, a structural transition is initiated and characterized by the formation of additional [Co···Co] linkages between neighboring $[CoC_4]_\infty$ ribbons leading to alternating short and long Co–Co···Co distances (Figure 1c: highlighted with dashed lines) [1,5]. This Peierls-type distortion is assumed to affect the superconducting transition below 4.5 K since it causes the electronic insulation of the $[CoC_4]_\infty$ ribbons [1,5]. Earlier studies also showed that the superconducting transition of $Sc_3CoC_4$ can be significantly influenced by only slight modifications of the electronic structure of the $[CoC_4]_\infty$ ribbons [5].

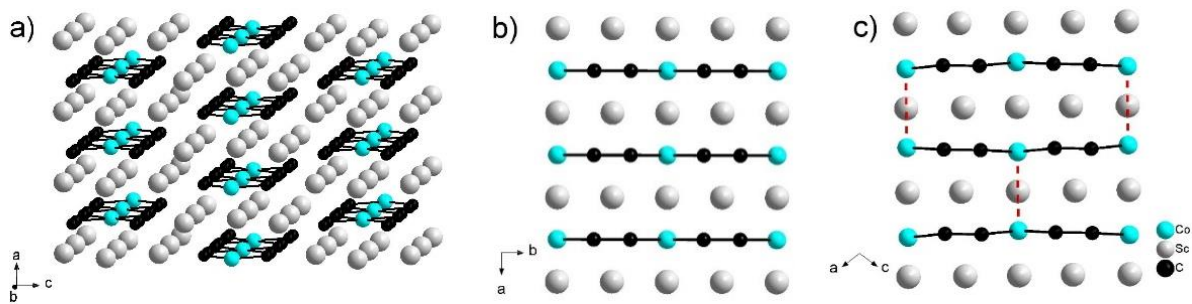

**Figure 1:** (Color online) Ball-and-stick representation of (a, b) the high temperature phase and (c) the low temperature phase of $Sc_3CoC_4$. The significant shortening of the [Co···Co] distances during the structural phase transition is highlighted by dashed lines in (c).



According to BCS theory,

$$k_B T_c = 1.13\, \hbar\omega_D\, exp[-1/N(E_F)V], \quad (1)$$

where $\omega_D$ is the Debye frequency, $k_B$ is the Boltzmann constant, $\hbar$ the reduced Planck constant and $V$ the electron-phonon interaction strength, the transition temperature $T_c$ depends on the electronic density of states at the Fermi level $N(E_F)$ [8,9]. Hence, suitable metal substitution inside the $[CoC_4]_\infty$ ribbons might provide a chemical control parameter of $T_c$. Indeed, stepwise substitution of Co by its neighboring elements in the solid solutions $Sc_3Co_{1-y}T_yC_4$ ($T$ = Fe, Ni; $y < 0.1$) causes a decrease of $T_c$ and a shift of the structural phase transition towards higher temperatures [5]. At higher substitution degrees ($y > 0.1$) both, the structural and the superconducting transition are even suppressed. It further follows from Eq. 1, that the electron phonon coupling might provide an additional control parameter for $T_c$. This can experimentally be probed by isotope substitution as according to BCS theory, the isotope exchange affects $T_c$ with the proportionality $M^{-\alpha}$ ($M$: isotope mass-ratio; $\alpha$: isotope coefficient). This follows directly from Eq. 1 employing the harmonic approximation for the phonon modes (from which consecutively the Debye frequency $\omega_D$ is obtained) and the Migdal adiabatic approximation [10] and by taking the ratio

$$\frac{T_c^{12C}}{T_c^{13C}} = \frac{1.13\hbar\omega_D^{12C} e^{-\frac{1}{N(E_F)V}}}{1.13\hbar\omega_D^{13C} e^{-\frac{1}{N(E_F)V}}} = \frac{\omega_D^{12C}}{\omega_D^{13C}} = \frac{\sqrt{\frac{k}{m^{12C}}}}{\sqrt{\frac{k}{m^{13C}}}} = \left(\frac{m^{12C}}{m^{13C}}\right)^{-\frac{1}{2}}, \quad (2)$$

thus resulting in a reference value of $\alpha=0.5$ [8,9]. Within this approach, $N(E_F)$ as well as the electron phonon coupling constant $\lambda$ and the force constant $k$ are independent of the atomic masses [11]. For most of the elemental superconductors $\alpha$ is found to be close to the BCS value of 0.5 (*e.g.* in Hg [12,13]). However, significant deviations from this idealized $\alpha$ value are discussed in the literature [14]: *e.g.* in elemental uranium ($\alpha$ = -2.2 [15,16]), in $H_3S$ at $p$=130 GPa ($\alpha$ = 2.37 [17]) or in $SrTiO_3$ ($\alpha \approx$ -10 [18]). Possible explanations for high positive values of $\alpha$ are the proximity of the superconducting state to a Lifshitz transition [19,20] and the breakdown of the Migdal approximation has previously been discussed as origin of the unusual $\alpha$ value of $H_3S$ under pressure [21]. Negative isotope coefficients (*inverse isotope effects*) have also been observed in the palladium hydrides PdH [22,23,24]. In these cases the



occurrence of negative $\alpha$ values has been attributed to an isotope mass dependency of the electron phonon coupling constant $\lambda$ and effects due to anharmonicity [25,26].

Theoretical studies suggest the superconductivity in $Sc_3CoC_4$ being mainly phonon-mediated involving couplings between the electronic structure to the lattice vibrations and librations of the $C_2$ dicarbido moieties within the $[CoC_4]_\infty$ ribbons [27]. In an attempt to provide experimental evidence for this theoretical prediction we performed a systematic $^{12}C/^{13}C$ labeling of the $[CoC_4]_\infty$ ribbons to investigate the carbon isotope effect on $T_c$.



**Experimental**

*Synthesis.* The samples of the substitution series Sc$_3$Co($^{12}$C$_{1-x}$$^{13}$C$_x$)$_4$ with $x$ = 0, 0.5, 1 [28] were synthesized under pure argon atmosphere by conventional arc melting of the pure elements (Sc: Smart Elements, 5N; Co: Cerametek Materials, 5N5; C) in the stoichiometric ratio 3:1:4 (Sc:Co:$^{12/13}$C) in a miniature furnace [29] installed in a glove box (argon inert gas). As $^{13}$C source, polycrystalline carbon powder (Sigma Aldrich, isotopic enrichment: 99 m% $^{13}$C) was compacted by spark plasma sintering (SPS) to a pellet prior to the synthesis employing the experimental conditions optimized by Villeroy *et al.* [30]. The $^{13}$C content of the $^{13}$C enriched pellet after SPS as specified by the supplier was checked by Raman and Simultaneous Thermal Analysis – Fourier Transform Infrared (STA-FTIR) spectroscopy. A $^{13}$C labeling degree close to 100% was confirmed by both characterization methods (see below). Magnetic measurements exclude the presence of any significant amounts of paramagnetic impurities (Fe or Ni) in the SPS compacted $^{13}$C pellet. As $^{12}$C source, graphite pellets (Alpha Aesar; 5N5, natural abundance of $^{13}$C: 1.1 at%) were employed. The stoichiometric parameter $x$ in the solid solution Sc$_3$Co($^{12}$C$_{1-x}$$^{13}$C$_x$)$_4$ has been adjusted by mixing both $^{12}$C and $^{13}$C sources in the appropriate ratio while the small natural abundance of the $^{13}$C isotope in the $^{12}$C source has been ignored. To ensure the highest possible sample homogeneity, all samples were flipped over and re-melted several times. The mass losses during the syntheses were less than 0.35 m%.

*Raman Spectroscopy.* In order to verify the $^{13}$C content of the $^{13}$C enriched pellet after spark plasma sintering (SPS), Raman spectra (Figure 2) of the $^{13}$C pellet and a regular $^{12}$C graphite pellet (containing the naturally abundance of the $^{13}$C isotope) were recorded with a *Thermo-Fischer* DXR Raman microscope equipped with a 10 mW Nd:YAG laser ($\lambda$ = 532 nm). For both samples, 6 independent bands of graphite (including combination bands and overtones) are present and assigned according to the literature [31,32]. The maximum positions of the peaks were used to determine the wavenumber of the independent bands and are listed in Table 1. A pronounced red-shift is present for all observed bands in case of the $^{13}$C enriched pellets compared to the regular $^{12}$C pellet due to the higher mass of the $^{13}$C isotope. All spectra do not show any satellite bands indicating relative pure isotope contents in all samples. In order to estimate the $^{13}$C isotope content in the enriched pellets, a simple harmonic oscillator model was applied where the frequency $v$ is $\propto \sqrt{\frac{k}{m}}$, with *m* being the isotope mass and the force constant



*k*. Similar to graphene, the same force constant *k* for both isotopes is assumed and this leads to a simple relation between the wavenumber and the isotope mass:

$$\nu(^{12.011}C)\sqrt{m(^{12.011}C)} = \nu(p \cdot {}^{13.00336}C)\sqrt{\left(m(^{12.011}C) + p \cdot \left(m(^{13.00336}C) - m(^{12.011}C)\right)\right)} \quad (3)$$

with *p* being the additional $^{13}$C content in comparison to the natural abundance in $^{12}$C [32]. Applying this relationship to all six independent Raman modes results in an averaged $^{13}$C enrichment of 103.4±3.7 % in the SPS-compacted pellet.

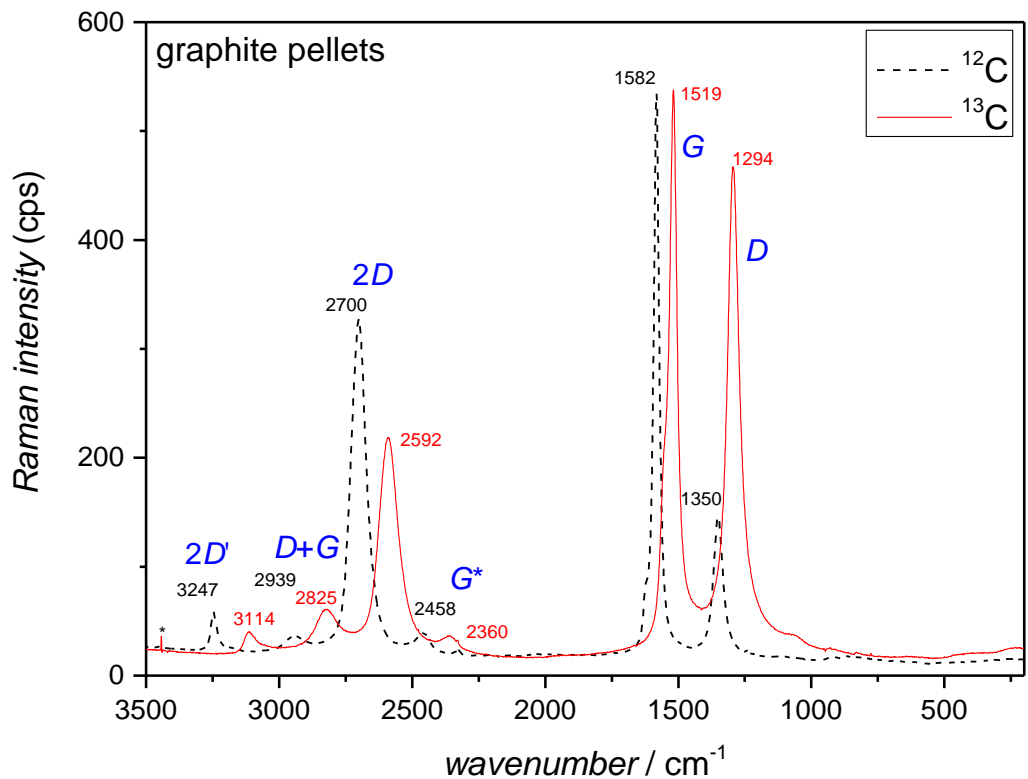

**Figure 2:** (Color online) Raman spectra of the $^{12}$C (broken black line) and the $^{13}$C enriched pellet (red solid line) after SPS-treatment. The observed bands are assigned to the typical graphite bands (blue capital letters) and the maxima of the peaks define the wavenumber of the individual bands.



| Graphite bands | D | G | G* | 2D | D+G | 2D' |
| --- | --- | --- | --- | --- | --- | --- |
| $\nu(^{12}C)$ / cm$^{-1}$ | 1350 | 1582 | 2458 | 2700 | 2939 | 3247 |
| $\nu(^{13}C)$ / cm$^{-1}$ | 1294 | 1519 | 2360 | 2592 | 2825 | 3114 |

**Table 1:** Assigned graphite bands of the $^{12}C$ and $^{13}C$ pellets. See Ref. [32] for the respective band assignments.

*STA-FTIR.* The SPS compacted $^{13}C$ pellet was characterized using the Perseus coupling of the *Bruker* Alpha FTIR (Fourier Transform Infrared) spectrometer mounted directly onto the gas outlet of the furnace of the STA (Simultaneous Thermal Analysis) *Netzsch* Jupiter F3. The standard DTA sample carrier with Alox crucibles was used to heat the carbon powder samples with 40 K/min between 35°C and 1200°C under an oxidizing atmosphere (synthetic air 50 ml/min, helium 20 ml/min). FTIR spectra were recorded continuously every 11s. In order to determine the $^{13}C$ enrichment in the $^{13}C$ pellet, the temperature dependent intensities at the characteristic asymmetric C=O stretching bands at 2265cm$^{-1}$ ($^{13}CO_2$) and 2365cm$^{-1}$ ($^{12}CO_2$) were analyzed. The $^{13}C$ enrichment of 99.2 m% is close to 100% and in line with the Raman results.

*Magnetization and resistivity measurements.* The magnetization of the solid solution series Sc$_3$Co($^{12}$C$_{1-x}$$^{13}$C$_x$)$_4$ with $x = 0, 0.5, 1$ was measured under Zero Field Cooled (ZFC) conditions right after synthesis under strict inert gas conditions between 2 K and 8 K in an external magnetic field of 0.5 mT. The earth magnetic field was compensated and the samples were cooled to 2 K using a commercial SQUID magnetometer (Quantum Designs *MPMS-7*). The uncertainty of the magnetic fields employed for the present studies was below 1 µT using the *TinyBee* setup [33]. High sample masses (at least 200 mg) were employed to study the superconducting transition temperature under ZFC conditions. Magnetization studies were performed at a temperature of $T = 2$ K at magnetic fields $\mu_0 H$ between -5 and 5 T with a *Quantum Design* MPMS-7 SQUID-magnetometer. Additionally, the electrical resistivity of single crystal whiskers with $x = 0$ and 1 were characterized between 1.8 and 300 K using a commercial *LR700* set-up in a *MPMS-7* (Quantum Design, for $x = 0$) and a *PPMS* (Quantum Design, $x = 1$).

*X-ray Diffraction.* Powder samples of the substitution series $Sc_3Co(^{12}C_{1-x}{}^{13}C_x)_4$ ($x$ = 0, 0.5, 1) were ground in a mortar. The powder was equally distributed on a flat sample holder and fixed by a small amount of grease. For all samples, powder diffraction patterns were subsequently collected using a *D5000* (Siemens, Bragg-Brentano geometry, Cu-$K_{\alpha 1}$ radiation,) equipped with a scintillation counter. Data was collected between $2\Theta = 5°$ and $80°$ in steps of $0.02°$. All observed reflections could be indexed based on the reported crystal structure of $Sc_3Co^{12}C_4$ (*i.e.* $x$ = 0) with the lattice parameters of the orthorhombic high temperature (HT) phase; $a$ = 3.3935(10) Å, $b$ = 4.3687(10) Å and $c$ = 11.9951(10) Å (space group *Immm*) [5]. No additional Bragg peaks hinting for any impurity phases could be identified. A *le Bail* fit of $Sc_3Co^{12}C_4$ ($x$ = 0) was performed employing the program *Jana2006* [34]. The cell parameters obtained ($a$ = 3.3954(10) Å, $b$ = 4.373(1) Å, $c$ = 11.993(4) Å) are again in good agreement with the previously reported values in the literature (see above) [5]. Single crystalline samples of $Sc_3Co(^{12}C_{1-x}{}^{13}C_x)_4$ ($x$ = 0, 1) were obtained by literature methods [3] and characterized at room temperature on a BRUKER D8 fixed-$\chi$ goniometer equipped with an INCOATEC Microfocus Source (Ag $K\alpha$ radiation). Precession images of reciprocal space showed no sign of twinning or other crystallographic issues in the samples.

*Solid-State NMR.* To investigate the presence of potential amorphous impurity phases which might cause the large shift of $T_c^{onset}$ observed with $^{13}C$ from a second manufacturer (Cambridge Isotopes), solid-state MAS NMR experiments were performed on a Bruker *Avance III* spectrometer operating at 7 T with a resonance frequency of 75.5 MHz for $^{13}C$ on native and labeled $Sc_3Co(^{12}C_{1-x}{}^{13}C_x)_4$ ($x$ = 0, 1) [35]. Due to the metallic nature of $Sc_3CoC_4$, the samples were diluted with freshly dried and calcined $SiO_2$ in a volume ratio of 1:1. Magic angle spinning was performed at 10 kHz employing a Bruker 4 mm MAS NMR probe and spectra were acquired employing $\pi/2$ pulse lengths of typically 5.5 µs, a relaxation delay of 3 s and accumulating 128 or 256 scans. The spin lattice relaxation time $T_1$ was determined to 0.52 s for $Sc_3Co^{13}C_4$. The chemical shift is referenced to TMS with adamantane as a secondary reference. The spectrum of the sample with $x$ = 1 is shown in Figure 3. The observed chemical shift for the dominant signal at 431 ppm is found to be in perfect agreement with the data published by Zhang *et al.* [36]. A minor additional signal is located at 447 ppm. We assign this signal to an impurity with an amount of approx. 2%. Increasing the relaxation delay to 120 s did not change the relative fraction of the impurity signal. From a simulation of the full spectrum, including spinning sidebands, the parameters for the CSA (chemical shift anisotropy) were obtained as $\delta_{iso}$ = 431 ppm, $d_{CSA}$ = 185 ppm, $\eta_{CSA}$ = 0.7.

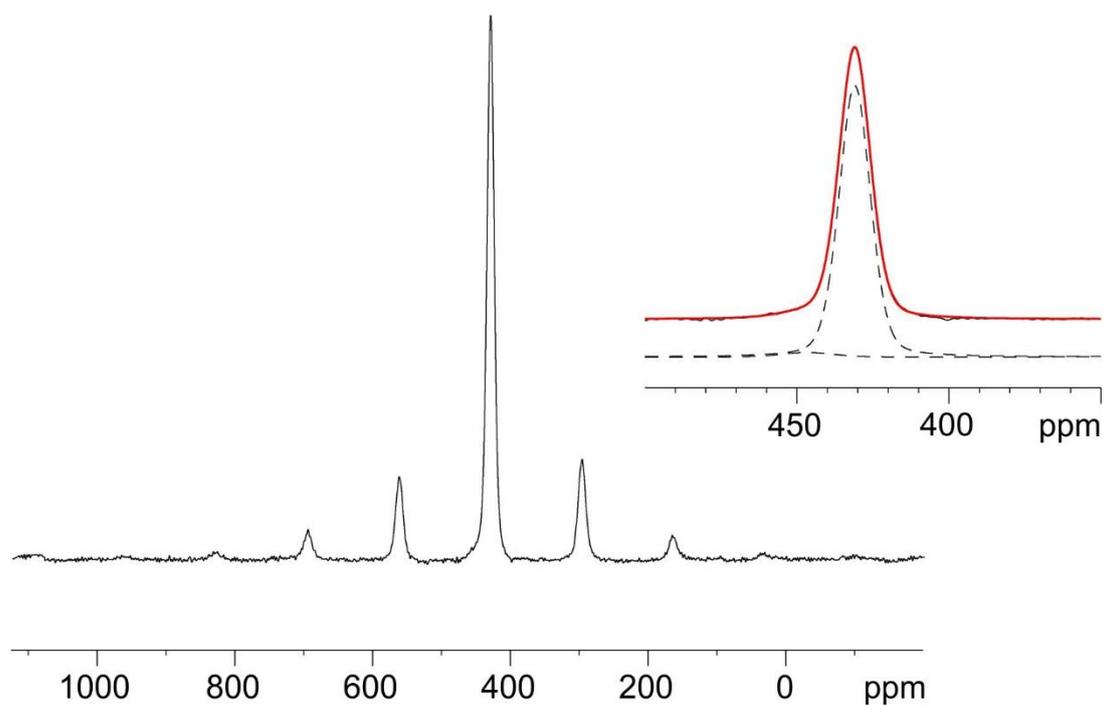

**Figure 3:** $^{13}$C-MAS-NMR spectra of Sc$_3$Co$^{13}$C$_4$ synthesized employing $^{13}$C purchased from *Cambridge Isotopes*. The inset shows a deconvolution of the center signal.

**Results and Discussion**

The temperature dependent volume susceptibility $\chi(T)$ of the polycrystalline solid solution series $Sc_3Co(^{12}C_{1-x}{}^{13}C_x)_4$ ($x$ = 0, 0.5, 1) [37] was measured between 2 K and 8 K (shown up to 6 K in Figure 4). From this data it can first of all be seen that in general the superconducting transition is characterized by a small superconducting volume fraction (*e.g.* for $x$ = 0: $\chi(2\ K)$ = 0.22%) and a rather broad transition region. This is in line with earlier findings for unsubstituted $Sc_3Co^{12}C_4$ (e.g. $\chi(2\ K)$ ~ 0.1% [38]); for further reports, see also references [1,2,3,39,40]. Similar characteristics have been observed for other quasi 1D superconductors, *e.g.* single crystals of $Tl_2Mo_6Se_6$ [41] and $Na_{2-\delta}Mo_6Se_6$ ($\chi$ < 0.1% with $\delta$ = 0.22-0.26 [42]). We note, however, that bulk superconductivity is observed for $Sc_3Co^{12}C_4$ below 1.55K (see Ref. [3]) and that the superconducting volume fraction increases significantly under external pressure ($\chi(2\ K)$ ~ 47% at 0.95 GPa) [38]. This resembles the behavior of the quasi-1D compound $NbSe_3$, being non-superconducting at ambient pressure, but becoming fully superconducting at 0.55 GPa [43].

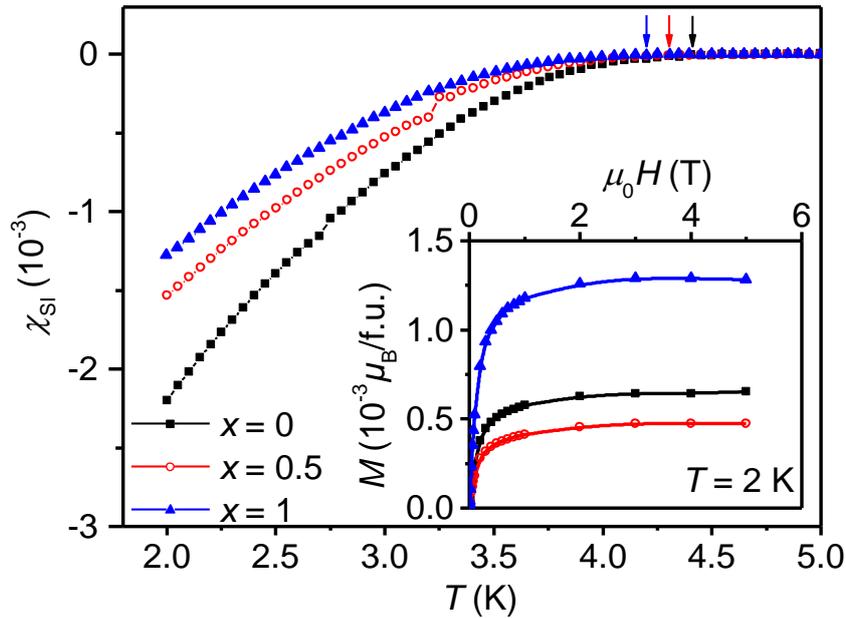

**Figure 4:** (Color online) Temperature dependent volume susceptibility $\chi(T)$ of the polycrystalline solid solutions $Sc_3Co(^{12}C_{1-x}{}^{13}C_x)_4$ ($x$ = 0, 0.5, 1). For a better comparison of the superconducting volume fractions, the $\chi(T)$ curves were normalized to each other relative to the according $\chi(8K)$ value. The arrows highlight the onset of the superconducting transition (for details see text). The inset shows the isothermal magnetization $M(\mu_0H)$ at 2 K after the subtraction of a linear paramagnetic contribution. The solid lines serve as a guide to the eye.

We further note, that the isothermal magnetization $M(\mu_0 H)$ of the substitution series $Sc_3Co(^{12}C_{1-x}{}^{13}C_x)_4$ at 2 K (<1.5m$\mu_B$/f.u. for all samples; see inset of Figure 4) is caused by spurious amounts of magnetic impurities (probably Fe and/or Ni in the $^{13}C$ precursor). Even though the amount of these impurities in the samples seems to follow no obvious trend upon successive $^{12}C/^{13}C$ substitution [5] we found, that $T_c$ (and thus the supposedly observed isotope effect) critically depends on the purity of the $^{13}C$ employed. For $Sc_3Co(^{12}C_{1-x}{}^{13}C_x)_4$ samples synthesized from $^{13}C$ powder purchased from *Cambridge Isotopes* we find isothermal magnetization values of up to 3.79 m$\mu_B$/f.u., resulting in a $T_c$ already as low as 2.65 K.

From the susceptibility measurements, the superconducting transition for each sample was graphically determined as the onset temperature, $T_c^{onset}$, at which the first variation of $\chi(T)$ from a linear regression within the experimental standard deviation is observed [44]. A decrease of $T_c^{onset}$ from 4.4(1) K ($x = 0$) to 4.2(1) K ($x = 1$, see Table 2) hints to a $^{13}C$ isotope shift very close the predicted value of the BCS theory, *i.e.* $T_c^{13C} = (13/12)^{-0.5} \times T_c^{12C} = 4.23$ K (where $T_c^{13C}$ and $T_c^{12C}$ are the predicted $T_c^{onset}$ values for fully substituted $Sc_3Co^{13}C_4$ and $Sc_3Co^{12}C_4$ samples, respectively). We also note that the superconducting volume fraction at 2 K further decreases with the decrease of $T_c^{onset}$.

To provide more experimental evidence for the observed isotope shift of $T_c^{onset}$ and in order to exclude any significant grain boundary effects on the results of the polycrystalline samples, we also performed temperature dependent electrical transport measurements, $\rho(T)$, on single crystal whiskers (sc) of $Sc_3Co(^{12}C_{1-x}{}^{13}C_x)_4$ ($x=0$ and 1; Figure 5). The whiskers were contacted by a four-point method and the values of $T_c^{onset}$ were graphically determined in the same way as for the $\chi(T)$ curves [44]. The decrease of $T_c^{onset}$ with increasing $^{13}C$ content from 4.4(1) K ($x = 0$) to 4.2(1) K ($x = 1$) is in line with our $\chi(T)$ studies of the polycrystalline samples (see Table 2) and yields an isotope coefficient of $\alpha_{\rho,sc} = 0.58$.

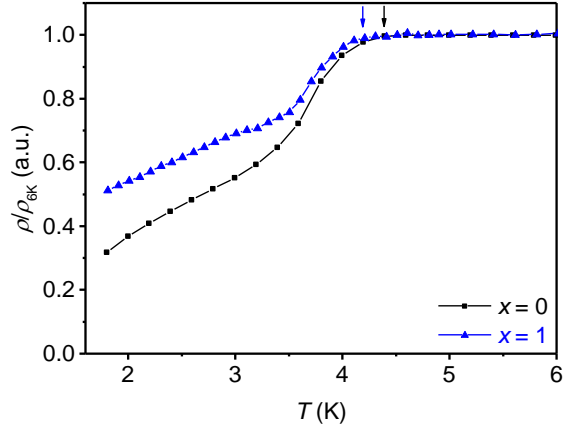

**Figure 5:** (Color online) Temperature dependence of the electrical resistivity, $\rho(T)$ of $Sc_3Co(^{12}C_{1-x}{}^{13}C_x)_4$ for $x = 0$ and 1; $T_c^{onset}$ is marked by arrows. The resistivity data has been normalized to $\rho(6K)$.

|  | 0 | 0.5 | 1 | $\alpha$ |
|---|---|---|---|---|
| $T_c^{onset}$ [K] $\chi_{pc}(T)$ | 4.4(1) | 4.3(1) | 4.2(1) | 0.58 |
| $T_c^{onset}$ [K] $\rho_{sc}(T)$ | 4.4(1) | - | 4.2(1) | 0.58 |

**Table 2:** Summary of $T_c^{onset}$ [K] and $\alpha$ values as determined from temperature dependent volume susceptibility $\chi(T)$ and electrical resistivity measurements $\rho(T)$ of polycrystalline (pc) and single crystalline samples (sc) with respect to the substitution degree $x$ in the substitution series $Sc_3Co(^{12}C_{1-x}{}^{13}C_x)_4$.

This observation supports the homogeneity of our polycrystalline samples in line with the findings of our powder/single crystal X-ray diffraction and solid state NMR studies. Again the superconducting transition occurs in a rather broad temperature range – in line with earlier studies of unsubstituted $Sc_3Co^{12}C_4$ samples [1,2,3,38] and other quasi 1D superconductors, *e.g.* $(SN)_x$ [2] and $Tl_2Mo_6Se_6$ [41,45,46]. Finite resistivity below $T_c^{onset}$ in our $\rho(T)$ data is another well-known feature of quasi 1D superconductors [47,48]. The broadness of the transition might be a consequence of a dimensional crossover, with 1D fluctuating superconductivity first occurring at higher temperatures causing a continuous downturn in the resistivity, while the transverse Josephson or proximity coupling among adjacent $[CoC_4]_\infty$ ribbons triggers a transition towards a 3D bulk coherent superconducting phase in the low temperature regime [3,41].

In summary, we have investigated the $^{12/13}$C isotope effect in the quasi-1D superconductor Sc$_3$Co($^{12}$C$_{1-x}$$^{13}$C$_x$)$_4$. The according isotope coefficient of $\alpha = 0.58$ obtained from our data is close to the ideal value of 0.5 predicted within BCS Theory. We note however, that because of the special characteristics of the $\chi(T)$ and $\rho(T)$ curves (broadness of the transition and change of the slope with $x$) due to the low-dimensionality of the system, the exact determination of the true isotope effect proves to be extremely delicate. In addition, we have also shown that the $T_c^{onset}$ value of Sc$_3$CoC$_4$ appears to depend critically on the amount of magnetic impurities contained in the samples. Using $^{13}$C precursors, which according to the certificate of analysis provided by the supplier contain approx. 750 ppm Fe, already reduce the $T_c^{onset}$ to 2.65K. Therefore, in the present study $^{13}$C powders of *Sigma-Aldrich* were used which did not show the presence of any significant magnetic impurities by detailed magnetic studies. However, also in case of this $^{13}$C source traces of Fe impurities (ca. 130 ppm according to the manufacturer) cannot be completely ruled out. This is due to the fact, that $^{13}$C carbon is usually produced by the catalytic reduction of $^{13}$CO with hydrogen using porous iron (via reduction of Fe$_2$O$_3$ by hydrogen) as a heterogeneous catalyst [49]. Hence, the obtained $^{13}$C powders systematically suffer from an inherent Fe contamination.

Accordingly, in the absence of ultra-pure $^{13}$C precursors lacking any traces of Fe impurities and in light of the above mentioned problems concerning the exact determination of $T_c^{onset}$ in such samples, the determination of the true $^{12}$C/$^{13}$C isotope effect of Sc$_3$CoC$_4$ remains an experimental challenge and warrants further exploration.

This work was supported by the Deutsche Forschungsgemeinschaft (DFG, German Research Foundation) - project number SCHE478/8-3 (SPP1178) and Grant No. JE 748/1, and by the Research Grants Council of Hong Kong Grants SBI17SC14 and IEG16SC03. We thank Prof. A. Kampf (University of Augsburg) for fruitful discussions, Prof. D. Volkmer, Dr. H. Bunzen and P. Beroll (University of Augsburg) for the performance of Raman measurements and Dr. R. Horny (University of Augsburg) for the STA-FTIR measurements of $^{12}$C and $^{13}$C labeled pellets.